\documentclass[aps, twocolumn, superscriptaddress, 
floatfix, letterpaper, preprintnumbers]{revtex4}
\usepackage{graphicx}

\newcommand{\be}{\begin{eqnarray}}
\newcommand{\ee}{\end{eqnarray}}
\newcommand{\el}{\nonumber \hfill \\}
\newcommand{\mat}{\left ( \begin{array}{cc}}
\newcommand{\emat}{\end{array} \right )}
\newcommand{\matf}{\left ( \begin{array}{cccc}}
\newcommand{\ematf}{\end{array} \right )}
\newcommand{\nn}{\nonumber }

\begin{document}

\preprint{SUNY-NTG-03/02}

\title{The QCD Phase Diagram at Nonzero Temperature, Baryon and
Isospin Chemical Potentials in Random Matrix Theory}
\author {B. Klein} 
\affiliation{Department of Physics and Astronomy, State University of New 
York at Stony Brook, Stony Brook, NY 11794-3800}
\author{D. Toublan}
\affiliation{Department of Physics, University of Illinois at 
Urbana-Champaign, Urbana, IL 61801-3080}
\author{J.J.M. Verbaarschot}
\affiliation{Department of Physics and Astronomy, State University of New 
York at Stony Brook, Stony Brook, NY 11794-3800}

\date{\today}

\begin{abstract}
We introduce a random matrix model with the symmetries of QCD  
at finite temperature and chemical potentials for
baryon number and isospin. 
We analyze the phase diagram of this model in the chemical potential
plane for different temperatures and quark masses.
We find a rich phase structure with 
five different phases separated by both
first  and second order lines. The phases are characterized
by the pion condensate and the chiral condensate  for {\it each} of the 
flavors. 
In agreement with lattice simulations, we find that in the phase with 
zero pion condensate the critical temperature depends in the same way
on the baryon number chemical potential and on the isospin chemical 
potential.
At nonzero quark mass, we remarkably find that the critical end point at 
nonzero
temperature and  baryon chemical potential is split in two by an
arbitrarily small isospin chemical potential. As a consequence, there
are {\it two} crossovers that separate the hadronic phase from 
the quark-gluon plasma phase at high temperature.
Detailed analytical results are obtained at zero temperature and
in the chiral limit. 
\end{abstract}
 
\pacs{}

\maketitle

\renewcommand{\theequation}{1.\arabic{equation}}
\setcounter{equation}{0}
\section{Introduction}
\label{section-1}
Currently, there is a strong interest in exploring the phase diagram 
of QCD at finite baryon density. A large number of possible phases
have been suggested for QCD at finite baryon 
density (see \cite{Rajagopal-Wilczek2,Hands:2001jn}
for a review). However, at this moment, the existence of any of these
phases has been confirmed  neither by first principle calculations nor
by the phenomenology of heavy ion collisions and neutron stars. 
Because of the phase of the fermion determinant,
standard Monte-Carlo simulations are only possible for small values
of the chemical potential \cite{fodor,owe,frith,maria,gupta,miya,gavai}. 
Neutron stars are probably the most likely 
candidates for high baryon density physics, but they are hard to observe 
and only few parameters can be measured accurately.  
Relativistic heavy ion collisions explore the region of low baryon density
and high temperature but give a complex picture of QCD at finite density.
However, an experimental observation of a tricritical point might be
within the realm of possibilities. Such a tricritical point was 
predicted on the basis of effective potentials \cite{gatto} and
simplified models such as random matrix
models \cite{shrock} and Nambu-type models \cite{berges}. 
Furthermore, in both neutron stars and relativistic heavy ion
collisions, the isospin density is different from zero. It is
therefore important both phenomenologically and theoretically to study
the influence of isospin on the phase diagram of QCD at nonzero baryon
chemical potential.

Our main goal is to study the phase diagram of QCD at nonzero temperature
and chemical potentials for baryon number and for isospin. 
The phase diagram of QCD with any of these external parameters equal
to zero has already been studied in a variety of ways. 
In particular, because lattice simulations are possible at 
nonzero isospin chemical
potential \cite{frank, kogiso}, the plane of zero
baryon chemical potential has been understood best.
At zero temperature, we expect a second order phase transition to a phase of 
condensed pions at an isospin chemical potential equal to half the pion mass.
Such type of transition occurs in any QCD-like theory with a chemical
potential for the charge of a Goldstone boson
\cite{kst,kstvz,TV,misha-son,domstrange,SST,SSSTV,sanvec,STV} 
\cite{STV2,kimvec,shins}.    
This prediction from effective Lagrangians 
has been confirmed by numerous lattice QCD simulations 
\cite{lat1,lat2,lat3,lat4,lat5,lat6,lat7,dom-kog,kogiso}.
~From lattice simulations \cite{dom-kog,kogiso} and a one-loop analysis of 
the effective
Lagrangian \cite{STV, STV2} it also follows that the second order line 
changes
into a first order line at a tricritical point with critical chemical potential
and temperature  on the scale of the pion mass.
At nonzero baryon chemical potential, analytical results have been
obtained for asymptotically large values of the chemical potential where
QCD can be analyzed perturbatively \cite{Bailin:bm, Son:1998uk}.
Lattice QCD simulations are only reliable for small values of the 
chemical potential \cite{maria, owe, fodor}. This leaves us with  
the bulk of the $\mu_B$-$T$-plane  which could only be analyzed in
simplified models such as Nambu
Jona-Lasinio models \cite{berges,krishna}, instanton liquid models
\cite{edward} and random matrix models \cite{shrock}.  

Random matrix models were introduced in the context of QCD to
describe the correlations of the low-lying eigenvalues of the QCD Dirac
operator \cite{SV,V}. It was shown that these models are equivalent to 
the 
mass term of a suitably chosen chiral Lagrangian which is determined
uniquely by the symmetries of the underlying microscopic 
theory \cite{OTV,DOTV,TV-correl,TV-beta}. Therefore, in the chiral limit,
chiral Random Matrix Theories provide an exact analytical description of
the low-lying Dirac spectrum.

In this article we use a Random Matrix Theory as a schematic model
for phase transitions in QCD at nonzero temperature and
chemical potentials. Such a model was first introduced in \cite{JV} to
describe the chiral phase transition in QCD at nonzero temperature.
Even more successful was the application of Random Matrix Theory
to QCD at nonzero baryon chemical potential. First, the failure of
the quenched approximation was explained analytically
\cite{misha}. Second, a
tricritical point was found in a model at nonzero baryon chemical 
potential
and temperature \cite{shrock}. Third, algorithms for QCD at finite
density could be investigated in detail \cite{adam, jun}. Fourth, 
the static part of effective Lagrangians for QCD with a chemical
potential for the charge of Goldstone bosons 
can be derived from a chiral random matrix model \cite{TV}.
In spite of the schematic nature of the random matrix model, we hope that
it will teach us more about the plethora of possible phases that may 
occur in QCD.

The organization of this paper is as follows. 
In section~\ref{section-2}, we deduce from general arguments the expected 
features of the QCD phase diagram for nonzero temperature, baryon and 
isospin chemical potentials.
In section~\ref{section-3}, we introduce our random 
matrix model. In section~\ref{section-4}, we derive an effective partition 
function 
in terms of the meson fields and reproduce the mean field results
obtained from a chiral Lagrangian. The phase diagram resulting from the 
random matrix model is obtained in section~\ref{section-5}. 
Concluding remarks are made in section~\ref{section-6}.

\renewcommand{\theequation}{2.\arabic{equation}}
\setcounter{equation}{0}
\section{QCD at nonzero chemical potentials and temperature }   
\label{section-2}

The QCD partition function at nonzero temperature and a chemical potential
for each quark flavor is given by
\be
\langle \prod_{f=1}^{N_f}\det( D + m_f + \mu_f \gamma_0) \rangle
\label{zqcd}
\ee
where the Euclidean Dirac operator is given by $D = 
\gamma_\mu(\partial_\mu+ i A_\mu)$ with $\gamma_\mu$ 
the Euclidean $\gamma$-matrices, and $A_\mu$ is 
an $SU(N_c)$ valued gauge potential. The quark masses are denoted by
$m_f$, and 
$\mu_f$ is the chemical potential for each flavor. The 
average is over the Euclidean Yang-Mills action. 
Below we mainly focus on QCD with two flavors and
nonzero baryon number and isospin chemical potential. 
In that case, 
the baryon number and isospin chemical potential are defined by
\be
\mu_B &=& \frac 12(\mu_1+\mu_2),\\
\mu_I &=& \frac 12 (\mu_1-\mu_2).
\ee
Before discussing the possible phases of
random matrix models with the symmetries of the QCD partition function,
we first make some general remarks on its phase diagram at nonzero
temperature, isospin and baryon chemical potentials. 

The case of $\mu_B=0$ and $m_\pi,\; T \ll \Lambda_{\rm QCD}$ can be
described in terms of a chiral Lagrangian. This Lagrangian has been
analyzed to one-loop order \cite{STV,STV2}. At low $T$, a second order
phase transition to
a pion condensation phase was found at $\mu_I = m_\pi/2$ where $m_\pi$
is the physical pion mass. 
For $\mu_I > m_\pi/2$ the chiral condensates rotate into a pion 
condensate,
denoted by $\rho$, and approach zero for 
$ \mu_I \gg m_\pi/2$.  Since a baryon chemical potential only excites states
for $\mu_B$ larger than the nucleon mass, we expect this second order line
to persist in the $\mu_I$-$\mu_B$-plane and to be parallel to the $\mu_B$
axis. 

The $\mu_I = 0$ plane was discussed
in detail in \cite{shrock}. We expect a region with broken chiral
symmetry (with chiral condensates $\langle \bar uu \rangle$ and
$\langle \bar d d \rangle $ both nonzero)
separated from the region of unbroken chiral symmetry by a
first order curve, denoted by $\mu_c(T)$, 
from the tricritical point to the $T=0$ axis and a 
second
order curve from the tricritical point to the $\mu_B = 0$ axis. 

For $\mu_I$ and $\mu_B$ both nonzero, 
there are eight possible phases with either of the chiral condensates, 
$\langle\bar uu\rangle$,
or $\langle\bar dd\rangle$, or the pion condensate 
$\rho=\frac{1}{2}(\langle \bar u \gamma_5 
d\rangle- \langle \bar d \gamma_5 u\rangle)$ equal to zero or not. 
Since the chemical potentials for the two flavors are different, there
is no reason to expect that $\langle \bar u u \rangle =
\langle \bar d d  \rangle$. 

In the limit $\mu_1 \gg \Lambda_{\rm QCD} $, one flavor decouples
and we are in a situation with only one flavor at nonzero
chemical potential. In this case, we
expect that $\langle \bar uu \rangle = 0$ and $\langle
\bar dd \rangle \ne 0$ for $\mu_2 < \mu_c(T)$ but vanishes across the
first order transition curve for $\mu_2 > \mu_c(T)$.

For $\mu_B =0$,  we have that $\mu_2 = -\mu_1$. Using that
\be
\det (D +m  - \mu_1\gamma_0) = {\det}^*(D +m +\mu_1\gamma_0),
\label{quenched-isospin}
\ee
we find that for equal quark masses the partition function (\ref{zqcd}) is
the phase quenched partition function for two flavors 
\cite{frank,misha-son}.

\renewcommand{\theequation}{3.\arabic{equation}}
\setcounter{equation}{0}
\section{Random Matrix Model}
\label{section-3}

In this article we study a random matrix model for QCD 
at nonzero chemical potentials and temperature. The idea is
to replace the matrix elements of the Dirac operator by Gaussian
random variables subject only to the global symmetries of the QCD
partition function. The dependence on the temperature and chemical
potentials enters through external fields structured according to these
symmetries. 

Our guiding principle for constructing a random matrix model
is dictated by the global symmetries of the QCD partition function. At zero
temperature and chemical potential, this amounts to replacing
the matrix elements of the Dirac operator by Gaussian random variables
subject to these global symmetries. 
The external field $\Omega(\mu_f ,T)$ representing
temperature and chemical potentials is introduced according to the
following criteria:
\begin{itemize}
\item The chemical potential breaks the global flavor symmetry in the
same way as in the QCD partition function.
\item The temperature field does not break global flavor symmetries.
\item For an anti-hermitian Dirac operator, the temperature field is
anti-hermitian, whereas the chemical potential field is hermitian.
\item The eigenvalues of the external field are $\mu_f \pm inT$. In 
this article we only consider the case $n= 1$.
\end{itemize}

For two flavors, the 
Dirac operator of the random matrix model is given by
\be \hspace*{-1.3cm}
\matf m_1 &\hspace*{-1.cm} \lambda &\hspace*{-1.cm} W+\omega(T)+\mu_1    
       &\hspace*{-1.cm}   0 \\
     -\lambda &\hspace*{-1.cm} m_2 &\hspace*{-1.cm}    0
&\hspace*{-1.cm} 
   W +\omega(T)+\mu_2 \\
   -W^\dagger+\omega(T)+\mu_1 &\hspace*{-1.cm} 0    
&\hspace*{-1.cm} m_1 
      &\hspace*{-1.cm} -\lambda  \\
       0  &\hspace*{-1.cm} -W^\dagger+\omega(T)+\mu_2 
&\hspace*{-1.cm} \lambda           &\hspace*{-1.cm} m_2 
\ematf .
\hspace*{-2cm} \nn \\
\label{3colD}
\ee 
We have also included a pion condensate source term, which in
QCD is given by
\be
i\lambda\bar \psi \gamma_5\tau_2 \psi,
\ee
where the Pauli matrix $\tau_2$ acts in flavor space.
The matrix elements of the $n \times n $ matrix $W$ are
complex 
with probability distribution given by
\be
P(W) = \exp( - n G^2 {\rm Tr} W W^\dagger).
\ee
The temperature field given by the matrix
\be
\omega(T) = \mat iT & 0 \\ 0 & -iT \emat
\ee
includes only the two lowest Matsubara frequencies.  At zero chemical
potentials, this temperature dependence will result in a second order
phase transition along the temperature axis \cite{JV}.
If we write the determinant as a Grassmann integral, the partition 
function
of our model is given by
\be
Z=\int {\cal D} W \prod_f d \psi^f d \bar \psi^f
 P(W) \exp[ -\sum_f\bar \psi^f D \psi^f  ] \nn \\ 
\label{zrandom}
\ee
where $D$ is the Dirac matrix given in (\ref{3colD}). In the thermodynamic
limit the partition function is a function of $m_1$, $m_2$, $\lambda$, $\mu_I$,
$\mu_B$ and $T$, but for brevity we will not display its arguments.

Other types of random matrix models have also been considered 
\cite{Benoit1, Benoit2, Benoit3, Pepin}. 
However, none of these models have been studied at nonzero
isospin chemical potential.
We mention models with a random gauge potential.
In that case, the matrix $W$ has the spin and color
structure of the usual Dirac operator,
\be
W \to i \sigma_\nu \tau_k A_\nu^k, \qquad \sigma_\nu = (-i, \sigma_k) 
\label{randomgauge}
\ee
but with $A_\mu^k$ a Gaussian $n\times n$ random matrix. Such types
of Dirac operators have the same spectral properties 
\cite{benoitspect} as the Dirac operator
in (\ref{zrandom}) and lead to a similar phase diagram.

\renewcommand{\theequation}{4.\arabic{equation}}
\setcounter{equation}{0}
\section{Effective Partition Function}
\label{section-4}
Because of the unitary invariance of the random matrix models, the 
partition
function can be rewritten in terms of invariant degrees of freedom only.
Below we rewrite the partition functions introduced in Section
\ref{section-3} in terms of these effective degrees of freedom.

We consider the random matrix model for QCD at nonzero temperature, 
baryon and isospin chemical potentials given in (\ref{zrandom}).
The Gaussian integration over the matrix 
elements of $W$ can be performed trivially. 
The resulting four-fermion interaction is
decoupled by means of a Hubbard-Stratonovich 
transformation at the expense of introducing mesonic degrees of freedom. 
After performing the Grassmann integrations, the partition function 
(\ref{zrandom})  
can thus be written as
\be
Z = \int {\cal D}A \exp(-{\cal L}(A, 
A^\dag)),
\label{za}
\ee
where
\be
{\cal L} &=& n G^2 {\rm Tr} (A-M^\dagger) (A^\dagger-M) - \frac n2 {\rm
Tr} \log Q' \nn 
\\
&=& n G^2 {\rm Tr} (A-M^\dagger)( A^\dagger-M) - \frac n2 {\rm Tr} \log
Q^\dagger Q,\nn \\
\label{zla}
\ee
and $A$ is an arbitrary complex $N_f\times N_f$ matrix. 
The determinant of the $4N_f\times 4 N_f$ matrix $Q'$, given by
\be\hspace*{-1.5cm}
 \left | \begin{array}{cccc}
A  &\hspace*{-1.cm} 0 &\hspace*{-1.cm} iT + \mu_B+\mu_I I_3 
 &\hspace*{-1.cm}0 \\
0 &\hspace*{-1.cm} A &\hspace*{-1.cm} 0 &\hspace*{-1.cm} 
-iT + \mu_B+\mu_I I_3 \\
iT+\mu_B +\mu_I I_3 &\hspace*{-1.cm} 0 &\hspace*{-1.cm} A^\dagger 
&\hspace*{-1.cm} 0 \\
0 &\hspace*{-1.cm} -iT +\mu_B+ \mu_I I_3&\hspace*{-1.cm} 0 &
\hspace*{-1.cm} A^\dagger 
\end{array} \right |,\hspace*{-2.5cm}\nn \\
\ee
with $I_3={\rm diag}(1,-1)$,
factors into  the determinant of
\be
Q = \left ( \begin{array}{cc}
A  &  iT + \mu_B+\mu_I I_3 \\
iT+\mu_B +\mu_I I_3 & A^\dagger  
\end{array} \right )
\ee
and its Hermitian conjugate. The mass matrix is given by
\be
M= \mat m_1 & -\lambda \\ \lambda & m_2 \emat .
\ee
By shifting $A$, 
we have absorbed the dependence on $M$ into the quadratic 
term.

At zero temperature and chemical potentials, the chiral random matrix
partition function is equivalent to the zero momentum part of 
the QCD chiral Lagrangian.
We will now show that this is also the case for the chiral
Lagrangian that can be derived from (\ref{za}). To derive this result, we
use the power counting scheme that is used in the construction of the
chiral Lagrangian:
$\mu_I$, $m_\pi\sim\sqrt{m}$ and $\sqrt{\lambda}$ are of the same
order \cite{kstvz}. 
We thus expand the Random Matrix Theory effective Lagrangian to first 
order 
in $m$ and $\lambda$ and second 
order in $\mu_I$ about the saddle point obtained for $\mu_I = m = 
\lambda=0$. 
The saddle point equation  given by
\be
\hspace*{-2cm}
G^2 ((AA^\dagger + T^2-\mu_B^2)^2 +4\mu_B^2 T^2 )A =
(AA^\dagger + T^2-\mu_B^2)  A \hspace*{-2cm}\nn \\
\ee
has two solutions,
\be
A = 0,
\ee
or the solution of
\be
G^2 ((AA^\dagger + T^2-\mu_B^2)^2 +4\mu_B^2 T^2 ) =(AA^\dagger +
T^2-\mu_B^2). \nn \\
\label{saddle}
\ee
For $\mu_I=0$ it is a natural assumption 
that the flavor symmetry is not spontaneously broken.
We can parameterize $A$ in the broken phase as
\be
A= \frac 1G \bar{\sigma}(\mu_B, T) \Sigma,
\label{apar}
\ee
where $\Sigma$ is a unitary matrix and
\be
\bar{\sigma}(\mu_B, T) =
\left ( \frac 12 + \frac 12 \sqrt{1-(4G^2\mu_B T)^2}
-T^2G^2 +\mu_B^2G^2 \right )^{1/2} \hspace*{-1.5cm}\nn \\
\ee
is a solution of the saddle point equation (\ref{saddle}).
Using the ansatz (\ref{apar}), the inverse of the matrix $Q$ for
$\mu_I =0$ is given 
by
\be
Q^{-1} = 
\frac 1{{\bar \sigma}^2/G^2-(\mu_B+iT)^2}
\left ( \begin{array}{cc} A^\dagger & -\mu_B -iT \\ -\mu_B -iT & A
\end{array} \right ). \nn\\
\ee
Inserting this result and the parameterization (\ref{apar}) 
in the chiral expansion of the Lagrangian (\ref{zla}), 
one easily derives
\be
{\cal L} &=& c_0(\mu_B, \mu_I, T) 
- n G\bar{\sigma}(\mu_B,T) 
{\rm Tr}(M\Sigma^\dagger +M^\dagger\Sigma) \nn\\
&&+ n \mu_I^2 G^2 \bar{\sigma}^2(\mu_B, T)c_2(\mu_B, T)
 {\rm Tr}( \Sigma^\dagger
I_3 \Sigma I_3 ),\nn\\
\label{massExp}
\ee
where $c_0(\mu_B, \mu_I, T)$ is independent of $\Sigma$ and 
\be
c_2(\mu_B, T)&=&\left( 1- \frac{(4 G^2 \mu_B T)^2}{4( 
\bar{\sigma}^2(\mu_B,T) -G^2(\mu_B^2-T^2))^2} \right).\nn \\
\ee
This effective Lagrangian coincides with the zero 
momentum part of the leading-order Chiral Lagrangian 
at zero temperature and baryon chemical potential derived in
\cite{TV,misha-son} based on the symmetries of QCD. A transition 
to a pion condensation phase takes place at $\mu_{I,c}^2 = 
m/(2 G \bar \sigma(\mu_B,T)c_2(\mu_B,T))$.  
The pion condensates vanishes for $\mu_I < \mu_{I,c}$.
For $\mu_I > \mu_{I,c}$ the chiral condensate rotates into
a pion condensate but the sum of the squares of the chiral condensate
and the pion condensate remains constant. 
The temperature and the baryon chemical potential 
affect both the magnitude and the orientation of the condensates.

\renewcommand{\theequation}{5.\arabic{equation}}
\setcounter{equation}{0}
\section {Phase Diagram }
\label{section-5}
As was argued in \cite{kst, kstvz}, the free energy
(\ref{massExp}) is completely determined by the transformation properties
of the QCD partition function. Since the random   matrix model has the
same global transformation properties as the QCD partition function, we
thus find the same low-energy limit. In this section we analyze the
random matrix partition function beyond this universal domain. In our
model we will be able to study the partition function to all orders in
the mass, chemical potential and temperature. We will show that the
inclusion of such non-perturbative contributions alters the nature
of the phase transition.

\subsection{Observables}

We consider three different observables, the chiral condensates
$\langle \bar u u \rangle$ and $\langle \bar d d
\rangle$, 
and the pion condensate 
$\frac{1}{2} (\langle \bar u \gamma_5 d\rangle- \langle \bar d \gamma_5 
u\rangle)$. 
They can be expressed
in terms of derivatives of the partition function,
\be
\langle \bar{u} u \rangle&=& \frac{1}{2n} \partial_{m_1} \log Z^{\rm eff} 
\nn \\
    &=&G^2 \left(\frac{1}{2}\langle A_{11}^* + A_{11} \rangle - m_1 \right),\\
\langle \bar{d} d \rangle &=&  \frac{1}{2n} \partial_{m_2} \log Z^{\rm 
eff} \nn \\
    &=&G^2 \left( \frac{1}{2}\langle A_{22}^* + A_{22} \rangle - m_2 \right),\\
\lefteqn{\frac{1}{2} (\langle \bar u \gamma_5 d\rangle- \langle \bar d 
\gamma_5 u\rangle)} \hspace{0.6cm}\nn \\
&=& \frac{1}{4n}\partial_{\lambda} 
\log Z^{\rm eff} \nn \\
&=&G^2 \left( \frac{1}{4}\langle A_{12} + 
A_{12}^* - A_{21}- A_{21}^* \rangle - 
\lambda \right).
\ee
The expectation values of the diagonal matrix elements of $A$ can
be interpreted as the chiral condensates, whereas its
off-diagonal elements represent the pion condensate.  

\subsection{Effective potential} 

In the large-$n$ limit, the partition function can be calculated by a 
saddle point approximation.  
To solve the saddle point equations,
we make an ansatz for the matrix $A$.
Since we have two independent chemical potentials,
the chiral condensates are not necessarily equal, whereas
for a sufficiently large isospin chemical potential we expect a 
pion condensate. Based on the expressions for the chiral condensate
and the pion condensate, we make the following ansatz for $A$
\be
A &=& \left( \begin{array}{cc} \sigma_1 & \rho \\ -\rho & \sigma_2 
\end{array} \right).   
\ee
In this parameterization, the different condensates are given by 
$\langle \bar u u \rangle =G^2(\sigma_1-m_1)$, $\langle \bar d d
\rangle=G^2( \sigma_2-m_2)$, 
and
$\frac{1}{2} (\langle \bar u \gamma_5 d\rangle- \langle \bar d \gamma_5 
u\rangle)=  G^2 (\rho - \lambda) $. 
Using this ansatz, we obtain the effective potential
\be
&&\frac{1}{n} {\cal L}  =  G^2 ((\sigma_1-m)^2 
+ (\sigma_2-m)^2 +2 (\rho-\lambda)^2 ) \el
&& -\frac{1}{2} \sum_{\pm} \log \left[ ((\sigma_1  +(\mu_1\pm 
iT))(\sigma_2  -(\mu_2\pm iT)) + \rho^2 )\right. \el
\hspace{0.1cm} &&\times\left. ( (\sigma_1  -(\mu_1\pm iT))(\sigma_2  
+(\mu_2 \pm iT)) + \rho^2 )\right].
\ee
\noindent From here on, we set $m_1=m_2=m$.
For $\rho=0$ this effective potential is a function of $\mu_f^2$. 
In particular, this means that its dependence on $\mu_I$ at $\mu_B= 0$
is the same as its dependence on $\mu_B$ at $\mu_I =0$. This implies
that for the critical temperature we have the relation
\cite{frith, kogiso} 
\be
\left. T_c(\mu_I)\right |_{\mu_B=0} = 
\left. T_c(\mu_B)\right |_{\mu_I=0}\quad {\rm for} \quad \rho =0.
\label{IBsymm}
\ee
The fermion determinant of the theory with $\mu_B= 0$ and equal quark 
masses
is equal to the fermion determinant of the phase quenched partition 
function.
We thus expect that the phase quenched approximation works in the phase where
the pion condensate vanishes.

To find the phase structure of our partition function,
we have to solve the saddle point equations 
\be
\frac{\partial {\cal L}}{\partial \sigma_1} = 0, \quad \frac{\partial 
{\cal L}}{\partial \sigma_2} = 0, \quad  \frac{\partial {\cal L}}{\partial 
\rho} = 0.
\ee
We will treat the following cases analytically: the chiral limit ($m=0$) 
at zero $T$ and at finite $T$, and the case $m  \neq 0$ at $T=0$. We will 
calculate the phase diagram for $T \neq 0$ and $m \neq 0$ by numerically 
minimizing the effective potential. In addition, some analytical results 
are obtained for large quark mass $m$.

\subsection{Chiral limit at $T=0$ } 
In the case of vanishing quark mass, zero diquark source, and zero
temperature, the effective  
potential simplifies to
\be
\frac{1}{n} {\cal L} &=&  G^2 (\sigma_1^2 + \sigma_2^2 +2 \rho^2 
) \el
&& -\frac12 \log \left( (\sigma_1 + \mu_1)(\sigma_2 - \mu_2) + 
\rho^2  \right )^2 \nn\\
&&
 - \frac12 \log\left( (\sigma_1 - \mu_1)(\sigma_2 +
\mu_2) + \rho^2 \right)^2.
\ee
In the chiral limit, chiral symmetry is broken spontaneously. As soon
as the isospin chemical potential is switched on, the chiral condensate
rotates into a pion condensate. Therefore, no 
phase exists where both condensates are nonzero. The saddle point equation
for $\rho$ has two possible solutions: $\rho = 0 $ and $ \rho \ne  0$. 
We first consider the case $\rho =0$.

For $\rho= 0$, the effective potential separates into the sum of free energies
for $\sigma_1$ and $\sigma_2$,
\be
\frac{1}{n}{\cal L} &=& \sum_{f=1,2} G^2 \sigma_f^2 - \frac 12 \log (\sigma_f^2
- \mu_f^2)^2.
\ee
The saddle point equations given by
\be
\sigma_f \left( G^2(\sigma_f^2 - \mu_f^2) -1 \right) = 0, \,\, f = 1, 2, 
\ee
have the solutions
\be
\sigma_f & = & 0, \,\, f =1, 2,\\
\sigma_f^2 &=& \frac{1}{G^2} +\mu_f^2\, ,\,\, f = 1, 2. 
\ee 
The contribution to the free energy from one flavor for these solutions is 
given by
\be
\Omega_f & = & -\log \mu_f^2,\\
\Omega_f &=& 1 + \log G^2 + \mu_f^2 G^2,
\ee
respectively. The full free energy is a sum over the contributions from 
both flavors. The two solutions are separated by a first order 
phase transition line where their free energy is equal,
\be
1+ \mu_f^2 G^2 +\log (\mu_f^2 G^2) &=& 0.
\label{mucrit}
\ee
The solution of this transcendental equation is given by
$\mu_f G = \mu_{c} G \approx 0.527697$ \cite{misha}. 
At this point a first order
phase transition to a phase with $\sigma_f =0$ takes place.
In the $\mu_1$-$\mu_2$-plane, we can thus distinguish four different 
phases with nonzero condensates in strips along the chemical potential
axes. The strips overlap in the center and form a region where both
chiral condensates are nonzero.

At zero quark mass, the Goldstone bosons are massless, and the 
critical value of the chemical potential for pion condensation is 
$\mu_I=0$. For $\mu_I > 0$, the chiral condensates rotate completely 
into a pion condensate, so that $\sigma_1=\sigma_2=0$.
The effective potential for $\rho$ is given by
\be
\frac 1n {\cal L} &=& 2G^2 \rho^2 - \log(\rho^2 - \mu_1\mu_2)^2.
\ee
The saddle point equation given by
\be
\rho(G^2(\rho^2-\mu_1\mu_2) -1) &=& 0
\ee
has again two solutions,
\be
\rho &=& 0, \el
\rho^2 &=& \frac{1}{G^2} +\mu_1\mu_2.
\ee 
A second order transition line is given by the hyperbola $\rho^2=0$ in the 
quadrants where $\mu_1\mu_2 < 0$.
The free energy of this phase is 
given by 
\be
\Omega_{\rho} &=& 2\left(1+\log G^2 + \mu_1 \mu_2 G^2 \right).
\ee

Finally, for $\mu_1=\mu_2$, the saddle point equations allow a solution
with $\rho \ne 0$ and $\sigma_1=\sigma_2 \ne 0$. However, the free
energy of this solution is higher than that of the solution with 
$\sigma_1=\sigma_2 \ne 0$ and $\rho = 0$. Physically this is clear,
since the pion condensate is expected to vanish for zero isospin
chemical potential.

\begin{figure}
\includegraphics[scale=0.60, angle=0]{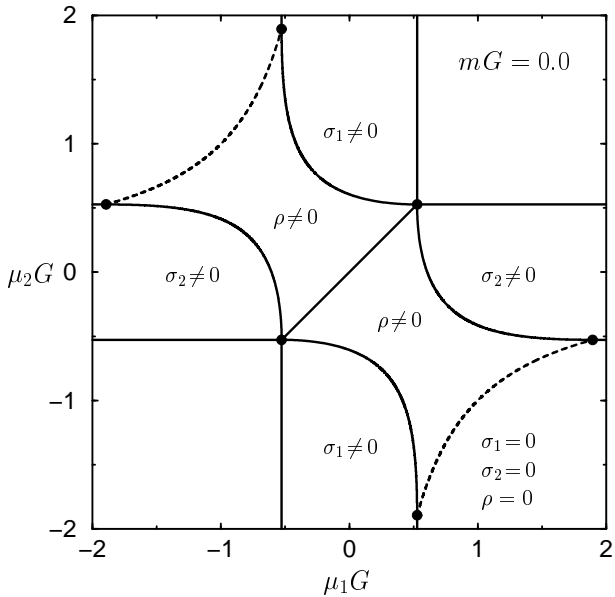}
\caption{Phase diagram of the RMT model for three colors in the chiral 
limit ($m=0$) at zero temperature. Solid (dashed) lines are lines of
first (second) order phase transitions.  Except for the four corner regions,
the different phases are marked by the nonvanishing condensate.
In the four corner  
regions we have that $\rho =\sigma_1=\sigma_2=0$.
 \label{fig1}} 
\end{figure}

Comparing the free energy of the pion condensation phase to those of the 
chiral condensation phases and the chiral restored phase, we can determine 
the remaining first order phase transition lines. The resulting phase 
diagram is shown in Fig.~\ref{fig1}.
We find a region of pion condensation in the center of the phase diagram. 
It is bounded by first order phase transitions towards phases with nonzero 
chiral condensate for one flavor, and by second order transition lines 
towards the chiral restored phase. The chiral condensation phases form the 
arms of a cross along the chemical potential axes.
Since $\sigma_1$ is independent of $\mu_2$, it is nonzero in a strip along 
the $\mu_2$-axis, and the same applies for the other flavor.  
The first order lines intersect in the two points $(\mu_1 G, \mu_2 G) 
\approx (\pm 0.527697, \pm 0.527697)$. The intersection points of the 
second order lines with the first order transitions between the pion and 
chiral condensation phases are at $(\mu_1 G, \mu_2 G)\approx (\pm 
0.527697, \mp 1.895025)$, and at the two points obtained by interchanging the 
values for $\mu_1$ and $\mu_2$.

\subsection{Chiral limit at $T\ne 0$ } 
At nonzero temperature, 
we analyze the phase structure in the same way as we did for $T=0$, 
and, initially, we also find the same phases. 

In a phase with vanishing pion condensate, the effective potential 
again separates into a sum over contributions of two different flavors,
\be
\frac{1}{n}{\cal L} =\sum_{f=1,2}G^2 \sigma_f^2 -\frac{1}{2} \log
((\sigma_f^2 - \mu_f^2 +T^2)^2 + 4 \mu_f^2 T^2) . \nn \\
\ee
This free energy was studied in \cite{shrock} for the case
of zero isospin chemical potential.
For each of the two flavors ($f=1, 2$), the saddle point equation  
\be
\lefteqn{ \sigma_f \Big(\sigma_f^4 - 2( \frac{1}{2G^2} +\mu_f^2 -T^2) 
\sigma_f^2 }\nn\\
 &+& \frac{\mu_f^2 -T^2}{G^2} +(\mu_f^2 +T^2)^2 \Big) =0  
\label{spsk}
\ee
has the solutions 
\be
\sigma_f & = & 0,  \\
\sigma_f^2 &=& \frac{1}{2G^2} + \mu_f^2 -T^2 \pm\frac{1}{2G^2} \sqrt{ 1 
-(4G^2 \mu_f T)^2}.\nn
\ee
The free energy for each flavor of these solutions is equal to
\be
\Omega_f &=& -\log(\mu_f^2+T^2),\\
\Omega_f &=& \frac 12+\log G^2 + G^2(\mu_f^2  -T^2) 
\pm\frac{1}{2}\sqrt{1-(4G^2 \mu_f T)^2} \el
& & -\frac{1}{2} \log\left(\frac{1}{2} \pm \frac{1}{2} \sqrt{1-(4G^2 \mu_f
T)^2}\right), 
\ee
respectively. For $(4G^2 \mu_f T)^2 < 1$,
the solution with the negative branch of the
square root will be discarded, since it does not minimize the free energy.
The second order phase transition line is
given by the condition that the two solutions coincide. One easily derives
\be
(\mu_f^2 -T^2) + G^2 (\mu_f^2 +T^2)^2=0. 
\ee
Since there is always the solution $\sigma_f=0$, there can be a first           
order transition when the coefficient of $\sigma_f^3$ in 
the saddle point equation (\ref{spsk}) becomes negative. A tricritical
point occurs where both the coefficient of $\sigma_f$ and $\sigma_f^3$
in (\ref{spsk}) vanish. This results in the equations
\be
\frac{1}{2G^2} + \mu_f^2 -T^2 &=&0, \el
\mu_f^2-T^2 + G^2 (\mu_f^2+T^2)^2 &=&0, 
\ee
with solution given by 
\be
\mu_{f, 3}^2 G^2 &=& \frac{\sqrt{2} -1}{4},  \quad f=1, 2, \el
T_3^2 G^2&=& \frac{\sqrt{2}+1}{4}  .
\label{tricrit}
\ee
Numerically, $T_3 G \approx 0.776887$, $\mu_{f, 3} G \approx 0.321797$,
which was also obtained in \cite{shrock}. 

\begin{figure*}
\includegraphics[scale=1.00, clip=, angle=0]{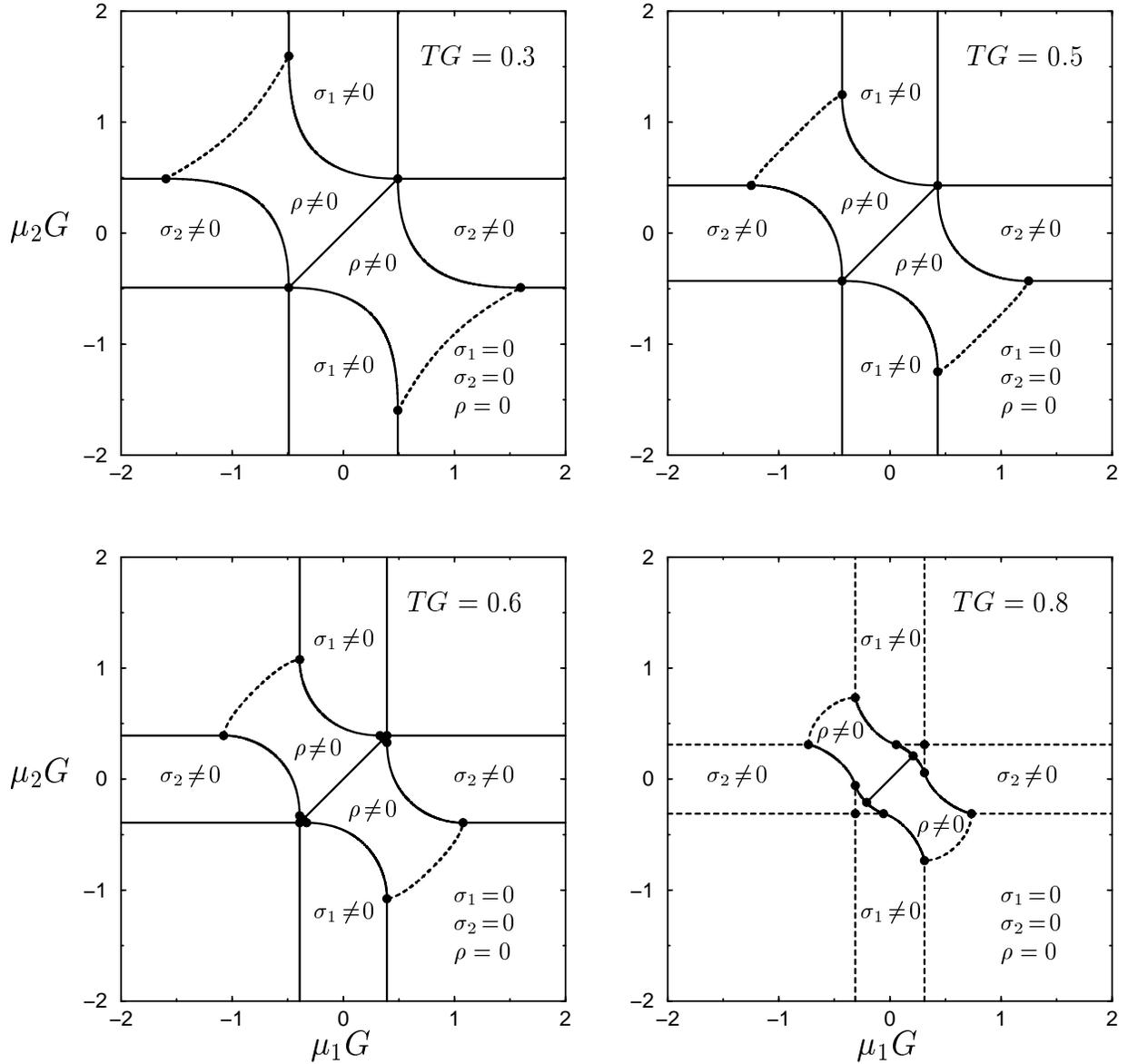}
\caption{Phase diagram of the RMT model  in the chiral
limit for different temperatures. Solid (dashed) lines are lines of first 
(second) order transitions. The values for the temperature are given in 
the figures. Except for the four corner regions, the 
phases are marked by condensates that do not vanish in 
the chiral limit.
  In the four corner regions we have that $\rho =\sigma_1=\sigma_2=0$.
For $TG= 0.6$,
the phase with $\sigma_1 \neq0$, $\sigma_2 \neq 0$ has emerged. For 
$TG=0.8$, the chiral restoration transitions have become second order 
transitions. The region in the center where both chiral condensation 
phases overlap and the chiral condensates for both flavors are nonzero
can be seen clearly. All condensates vanish at $TG=1$. 
\label{fig2}}
\end{figure*}  

In Fig.~\ref{fig2}, we show the phase diagram for fixed temperature
in the 
$\mu_1$-$\mu_2$ chemical potential plane. The transition lines for each 
flavor are straight and constant in the chemical potential for the other 
flavor. In this plane, the phase transition lines change in their entirety 
from 
first to second order when we pass the tricritical temperature
(\ref{tricrit}) from 
below. The critical chemical potential of the transition 
becomes smaller with increasing temperature. A baryon
chemical potential that is large enough will destroy the pion condensate. 

Since the two chiral condensates are independent of one another and depend 
only on the chemical potential for the respective flavor, we again have 
four phases. A phase  where chiral symmetry is restored
($\sigma_1=\sigma_2=0$), and phases where 
either one  or both of the chiral condensates are nonzero. The free 
energies are simply given by the sum of the one-flavor free energies.

At nonzero isospin chemical potential, we expect that in the limit of
massless quarks the chiral condensates are completely rotated into 
a pion condensate. The effective potential for vanishing $\sigma_f$
and nonzero pion condensate $\rho$ becomes
\be
\frac{1}{n}{\cal L} &=& 2 G^2 \rho^2 - \log \left((\rho^2 - \mu_1
  \mu_2 +T^2)^2  + T^2 (\mu_1+\mu_2)^2  \right). \nn \\
\ee
The saddle point equation reads
\be
&&G^2 \rho \Big( \rho^4 - 2(\frac{1}{2G^2} +\mu_1\mu_2 -T^2 )\rho^2 
+\frac{1}{G^2} (\mu_1\mu_2 -T^2)  \nn\\&& +(\mu_1\mu_2
-T^2)^2 + T^2 (\mu_1+\mu_2)^2 \Big)=0.
\label{saddlerho}
\ee 
It has the solutions
\be
\rho&=&0, \\
\rho^2 &=& \frac{1}{2G^2} +\mu_1\mu_2 -T^2 \pm \frac{1}{2G^2}
\sqrt{1-(2 G^2  (\mu_1+\mu_2) T)^2}. \nn
\ee 
Again the solution with the negative branch of the square root has a larger
free energy. 
A line of second order phase transitions is determined by the condition 
that the two solutions coincide:
\be
\hspace*{-1cm}
(\mu_1\mu_2 -T^2) + G^2(\mu_1\mu_2 -T^2)^2 + G^2 T^2 (\mu_1+\mu_2)^2=0. 
\hspace*{-2cm}
\nn \\
\ee
A first order phase transition may occur when the coefficient of
$\rho^3$ in (\ref{saddlerho}) vanishes. However, we will see below
that this happens in a region where solutions with nonzero chiral
condensate have a lower free energy.

In Fig.~\ref{fig2} we show the phase diagram in the 
$\mu_1$-$\mu_2$-plane for zero quark mass and temperatures equal to
 $TG=0.3$, $TG=0.5$, $TG=0.6$ and $TG= 0.8$. The  first order
lines in the phase diagram
are  obtained by combining the results for the 
free energies of the phases discussed above.

The phase diagram at $T=0$ has been described in the previous section. 
With increasing temperature, the second order transition between the pion 
condensation phase and the chiral restored phase moves towards the origin. 
The effect of the temperature term on the phase diagram is a shrinking of 
the condensate phases. The critical chemical potential for the chiral 
restoration transition decreases with temperature, so that the transition 
lines move toward the axes as well. 

At lower temperatures, the phase where both chiral condensates are nonzero 
simultaneously is always higher in energy than the pion condensation 
phase, 
and consequently it is not realized. At a temperature of $T G \approx 
0.548047$, a first order transition between the pion condensation phase 
and the phase with $\sigma_1 \neq 0$, $\sigma_2 \neq 0$, and $\rho=0$ 
emerges, and it 
appears at the intersection points of the first order transitions, 
$\mu_1 G=\mu_2 G \approx 0.413485$, around the line $\mu_1=\mu_2$. The 
upper boundaries of this phase are always the transition lines where 
either of the two chiral condensates vanish.

The position of the tricritical point in the $\mu_f$-$T$-plane for any of 
the two chiral condensates $\sigma_f$ is unaffected by the presence of the 
second chemical potential. Therefore, we find that at a temperature 
$T_3 G = \frac{1}{2} \sqrt{\sqrt{2}+1} \approx 0.776887$, 
the phase transition 
lines between the chiral condensed phases and the chiral restored phase in 
the chemical potential plane become second order transition lines in their 
entirety. 

Above this temperature, the regions with nonzero chiral condensates along 
the $\mu_1$- and $\mu_2$-axes are bounded by second order transition lines 
to the chiral restored phase. We can observe the region where both phases 
with chiral condensation overlap. The pion condensation phase in the 
center is separated from the chiral condensation phases by first order 
transition lines, and from the chiral restored phase by a second order 
transition line. 

All condensation phases vanish at the same temperature $T G =1$.

\subsection{Zero temperature limit at nonzero quark mass}
 
Away from the chiral limit, we can only solve the saddle point equation 
analytically for $T=0$. The results of this particular case
 will be discussed in this section. 
The phase diagram at nonzero quark mass is qualitatively different
from the phase diagram in the chiral limit. First, 
the chiral condensates, $\sigma_1$ and $\sigma_2$, are 
 not good order parameters anymore, and,
second, we expect a phase transition to a phase with nonzero pion 
condensate for $\mu_I = m_\pi/2$. However, in this case the chiral
condensates are non-vanishing in the phase with $\rho\ne 0$.

The effective potential is given by
\be
\frac{1}{n} {\cal L} & = & G^2 ((\sigma_1-m)^2 + (\sigma_2-m)^2 +2 
\rho^2) \el
& &      - \frac12 \log\left( (\sigma_1  +\mu_1)(\sigma_2  -\mu_2) +
  \rho^2 \right)^2 \el 
& &      -\frac12 \log \left( (\sigma_1  -\mu_1)(\sigma_2  +\mu_2) +
  \rho^2 \right)^2. 
\ee
Obviously, it is symmetric under a simultaneous interchange of the chiral 
condensates $\sigma_1$ and $\sigma_2$ and the two chemical potentials 
$\mu_1$ and $\mu_2$. The saddle point equations are given by
\be
 2G^2(\sigma_1-m) &=& \frac{\sigma_2-\mu_2}
{(\sigma_1+\mu_1)(\sigma_2 - \mu_2)+\rho^2} \el 
&&+\frac{\sigma_2+\mu_2}     
{(\sigma_1-\mu_1)(\sigma_2 + \mu_2)+\rho^2} ,\\
 2G^2(\sigma_2 -m)&=& \frac{\sigma_1+\mu_1}
{(\sigma_1+\mu_1)(\sigma_2 - \mu_2)+\rho^2} \el 
&&+\frac{\sigma_1-\mu_1}     
{(\sigma_1-\mu_1)(\sigma_2 + \mu_2)+\rho^2} ,\\
 2G^2\rho &=& \frac{\rho}
{(\sigma_1+\mu_1)(\sigma_2 - \mu_2)+\rho^2} \el 
&&+\frac{\rho}     
{(\sigma_1-\mu_1)(\sigma_2 + \mu_2)+\rho^2} .
\label{sp3}
\ee
As we see from the last equation in (\ref{sp3}), there is always a 
solution 
with $\rho = 0$. We expect that this is the actual minimum of the free 
energy below the critical chemical isospin potential for pion 
condensation. In this case, as for $m=0$, the saddle point equations
for the two chiral condensates decouple and are given by
\be
G^2 (\sigma_f-m)(\sigma_f^2 -\mu_f^2) -\sigma_f =0. 
\ee
Although this third-order equation can be solved analytically, it is
more instructive 
to expand it in powers of $m$. To first order in $m$, we find for 
$\sigma_f$, $f = 1, \, 2$,
\be
\sigma_f &=& \pm\frac{1}{G} \left( \sqrt{1 + \mu_f^2 G^2} +\frac{mG}{2
    (1+\mu^2_f G^2 )}  + 
{\cal O}(m^2 G^2)\right) \nn \\
         &=& \pm\frac{1}{G} \left( 1 + \frac{1}{2} \mu_f^2 G^2 +\frac{1}{2} 
mG + {\cal O}(m^2G^2, \mu^4_fG^4)\right)\!, \label{spm1} \nn \\
\\
\sigma_f &=& \frac{\mu_f^2 G^2}{1+\mu_f^2 G^2} m + {\cal
O}(m^3G^2). \label{spmm}
\ee
The free energy of these solutions is given by
\be
\label{5.39}
\Omega_f &=& 1 + \log G^2 +\mu_f^2 G^2 - 2mG  \\
         & &  +{\cal O}(m^2G^2), \nn \\
\Omega_f &=& - \log(\mu_f^2) - {\cal O}(m^2 G^2),
\ee
respectively.
The solutions (\ref{spm1}) minimize the free energy for small values of
the chemical potential, and the solution 
(\ref{spmm}) minimizes the free energy 
for large values of the chemical potential. 
As in the chiral limit, we once again have four phases, where either 
chiral symmetry is broken spontaneously and the chiral order parameters 
$G \sigma_f$ are of 
${\cal O}(1)$, or where it is broken only explicitly and the chiral 
order parameters are of the order of the quark mass, ${\cal O}(m)$.
The free energies in these cases are given by the sum of the free
energies for each flavor.
In contrast to the case of the chiral limit $m=0$, the phase with 
$\sigma_1 \neq 0$, $\sigma_2 \neq 0$ appears at the center of the phase 
diagram. 

By matching the free energies for the phases with large and small values 
of the chiral condensates, we obtain the correction to the critical 
chemical potential due to the finite quark mass $m$. The critical chemical 
potential shifts to
\be
\mu_{f, c}^\prime G &=& \mu_{c} G \left(1+ \frac{mG}{1+\mu_{c}^2G^2} + 
{\cal O}(m^2G^2)\right)\!,
\label{5.41}
\ee
where $\mu_{c} G$ is the result for the chiral limit, obtained in 
(\ref{mucrit}).
 
For $m \neq 0$, both the pion condensate as well as the 
chiral condensates are nonzero in the phase with $\rho\ne 0$. 
We have to solve the full system of three saddle point equations. In
fact, in this case the analytical solution is relatively simple.
The two chiral condensates are related by the equation
\be
\sigma_1 - \sigma_2 = m \frac{ \mu_1 + \mu_2}{ \mu_1 - \mu_2}.
\ee
The solution for $\sigma_f$, $f= 1, 2$,  is given by  
\be
\sigma_f  = m \mu_f \frac{ \mu_1 + \mu_2}{(\mu_1 - \mu_2)^2}  
+ \frac{2m}{G^2} \frac{1}{(\mu_1-\mu_2)^2 -4m^2}. 
\label{fullssp} 
\ee
The pion condensate then follows from
\be
(\sigma_1-m)( \sigma_2-m) + \rho^2 &=& \frac{1}{G^2} +\mu_1 \mu_2
-4\frac{m^2 \mu_1 \mu_2}{ (\mu_1 - \mu_2)^2}.\nn \\
\label{5.44}
\ee
The free energy of this solution is
\be
\Omega(m, \mu_1, \mu_2) &=& 2\left(1+ \log G^2 + \mu_1\mu_2 G^2 
+m^2 G^2 \right )\el  & &
-m^2 G^2\frac{(\mu_1+\mu_2)^2}{(\mu_1 -\mu_2)^2} 
\el
& & -\frac 12 \log\left( \frac{(\mu_1-\mu_2)^2}{ 
(\mu_1-\mu_2)^2 -4m^2}\right)^2. 
\label{5.45}
\ee
For nonzero quark mass $m$, the phase in 
which $\rho \neq 0$ does not extend to zero isospin chemical potential. 
For $\mu_B =0$, the onset chemical potential follows by equating
the free energy $\Omega_1+\Omega_2$ of (\ref{5.39}) to the free energy
of the pion condensed phase given in eq. (\ref{5.45}). It is given by
$\mu_{I, c}^2=\frac{m}{2G}+{\mathcal O}(m^2)$
which identifies $\sqrt{2m/G}$ as the pion mass for $\mu_B = T = 0$.

\begin{figure}
\includegraphics[scale=0.60, clip=, angle=0]{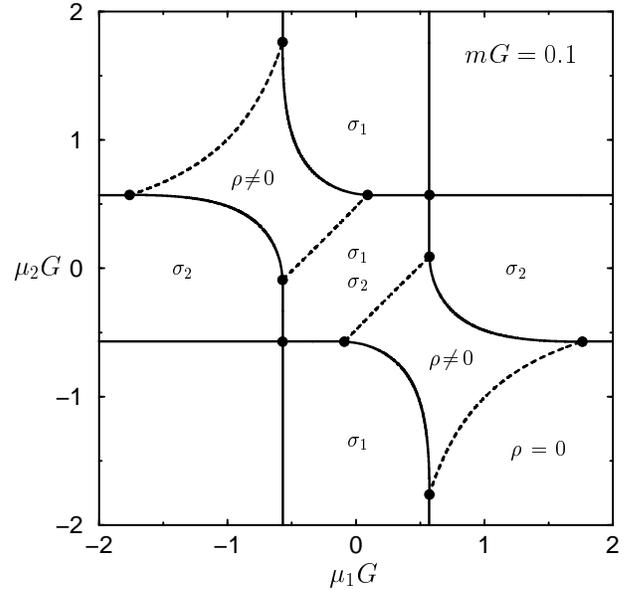}
\caption{Phase diagram of the RMT model for three colors at a value of the 
quark mass of $mG=0.1$. Solid (dashed) lines are lines of first (second) 
order transitions.
Except for the four corner regions,
the different phases are marked by 
the condensate that does not vanish in the chiral limit. The condensates
that are not displayed are of ${\mathcal O}(m)$.
In the four corner regions we have that $\rho =0$,
$\sigma_1 = {\mathcal O}(m)$ and $\sigma_2 = {\mathcal O}(m)$. }
\label{fig3}
\end{figure}

By putting $\rho^2=0$ in eq. (\ref{5.44}), we obtain the complete second order 
transition line that bounds the pion condensation phase at low as well as 
high isospin chemical potential. Parametrized in terms of the baryon and 
isospin chemical potentials, it is given by
\be
&&\left( \frac{\mu_B^2-\mu_I^2}{4 \mu_I^2} + \frac{1}{2G^2} 
\frac{m}{\mu_I^2-m^2} \right) m^2 \frac{\mu_B^2}{\mu_I^2}
-\mu_B^2 +\mu_I^2 \el
&&-\frac{1}{G^2} \frac{\mu_I^2}{\mu_I^2 - m^2} + 
\frac{m^2}{4 G^4}\frac{1}{(\mu_I^2-m^2)^2} =0. 
\ee
For $\mu_B= 0$ we again find a critical chemical potential given by
$\mu_{I, c} = m/2G + {\mathcal O}(m^2)$.
The phase diagram in the $\mu_1$-$\mu_2$-plane for $mG=0.1$ and zero
temperature is shown in Fig.~\ref{fig3}. The
qualitative difference to the massless case is the appearance of
a region where both chiral condensates are nonzero in the center of the
phase diagram. The dashed lines that border this region cross the 
$\mu_I$-axis ($\mu_B=0$) at $\mu_I = \pm m_\pi/2$ and are roughly constant 
in $\mu_B$. 
They coincide in the chiral limit, and the central region becomes a phase 
of
nonzero pion condensate and zero chiral condensates (see
Fig.~\ref{fig1}). For small quark masses $m$, the 
phase diagram at large values of the chemical potentials is almost 
unchanged in comparison to the phase diagram we found in the chiral limit. 
The transition lines are only shifted by small corrections of the order of 
the mass $mG$.

\begin{figure*}
\includegraphics[scale=1.00, angle=0]{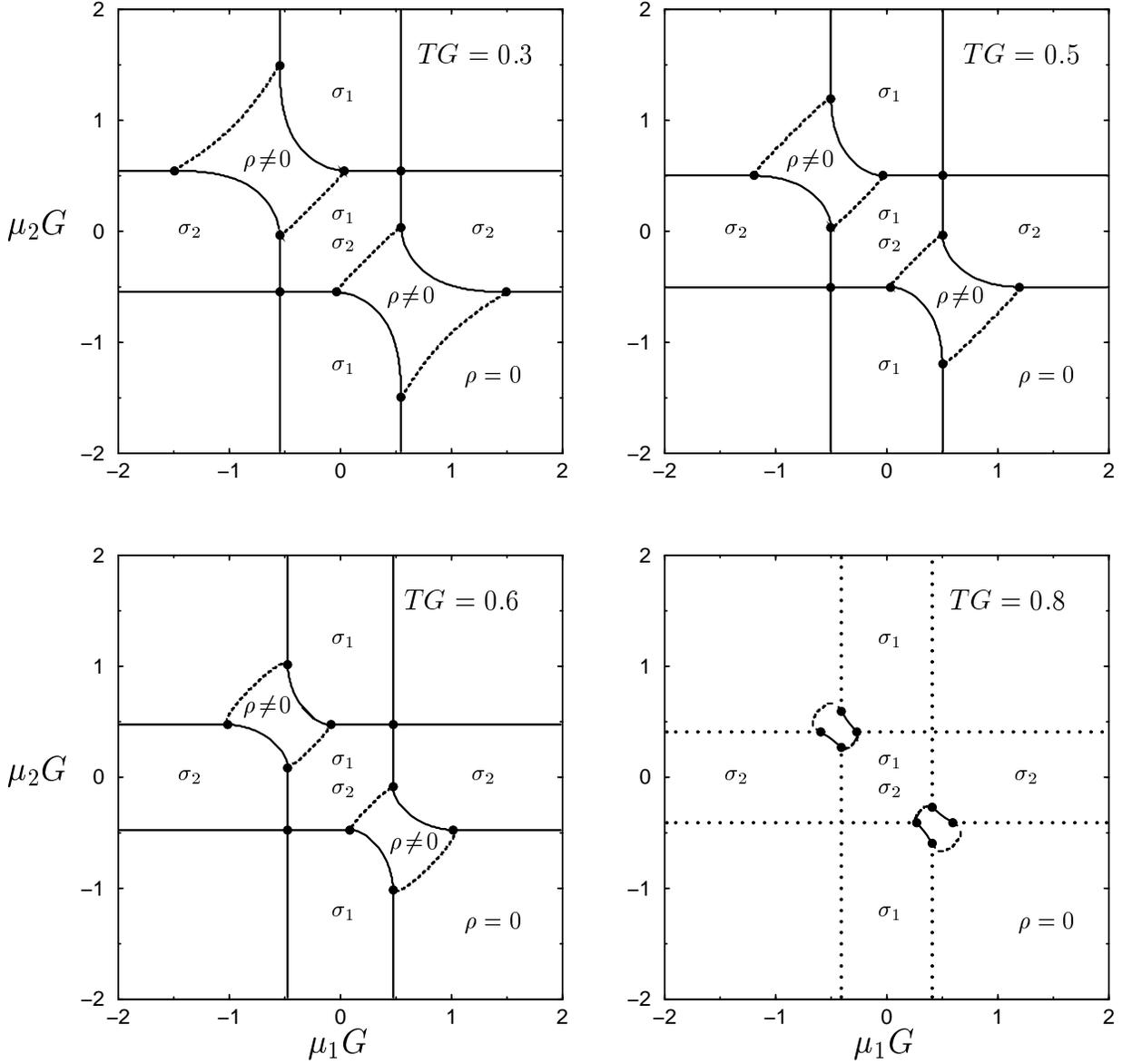}
\caption{Phase diagram of the RMT model for three colors for a value of
the quark mass of $mG=0.1$. Values for the temperature are given in the 
figures. Solid (dashed) lines are lines of first (second) 
order transitions. Above the critical temperature, the chiral restoration 
transition becomes a crossover, denoted by dotted lines. 
Except for the four corner regions,
the different phases are marked by  
the condensate that does not vanish in the chiral limit. The chiral condensates
are not displayed when they are of ${\mathcal O}(m)$.
In the four corner regions we have that $\rho =0$,
$\sigma_1 = {\mathcal O}(m)$ and $\sigma_2 = {\mathcal O}(m)$.
 \label{fig4}} 
\end{figure*}
\begin{figure}[h!]
\includegraphics[scale=0.40, angle=0, draft=false]{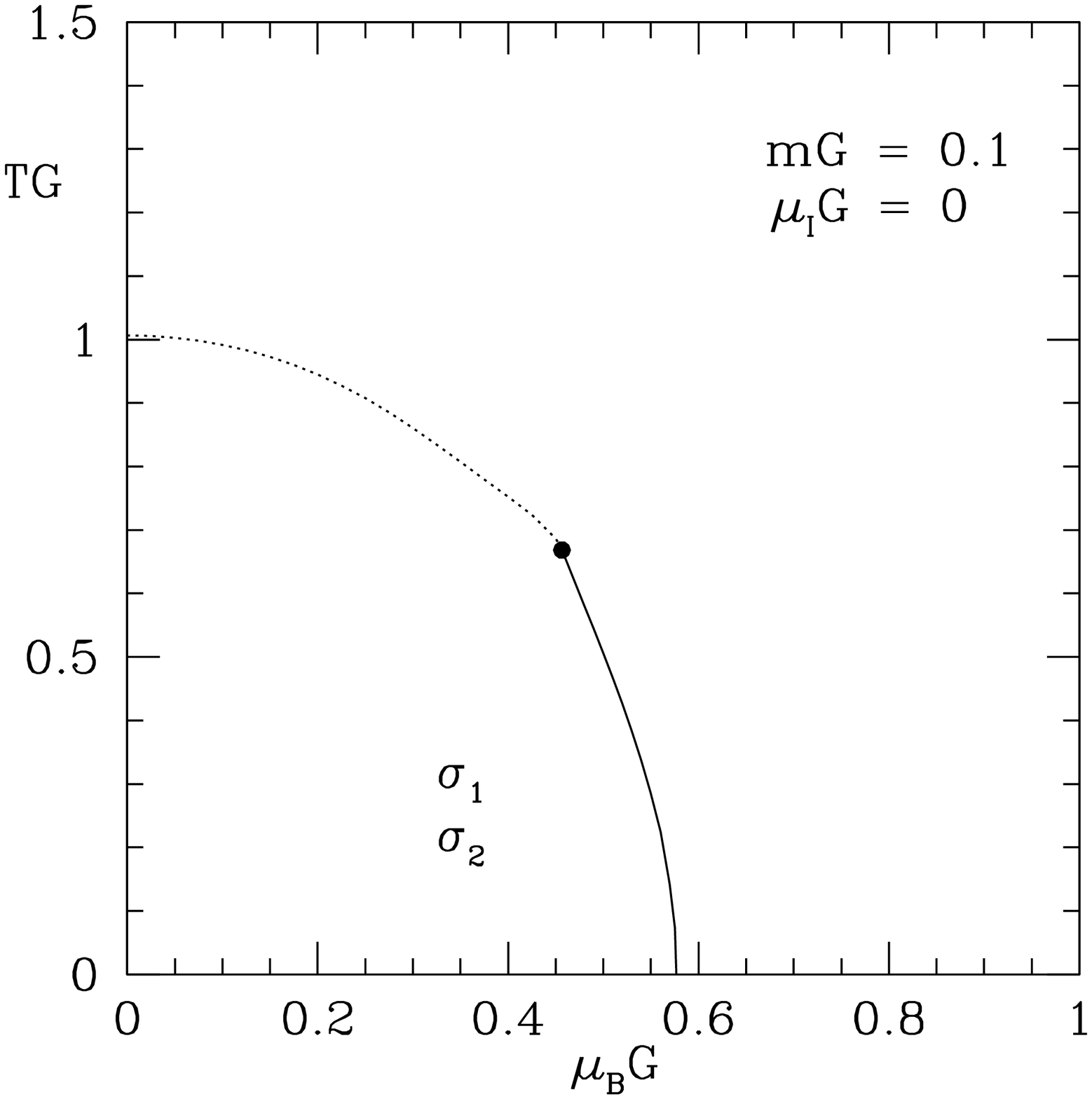}
\vskip 1cm
\includegraphics[scale=0.40, angle=0, draft=false]{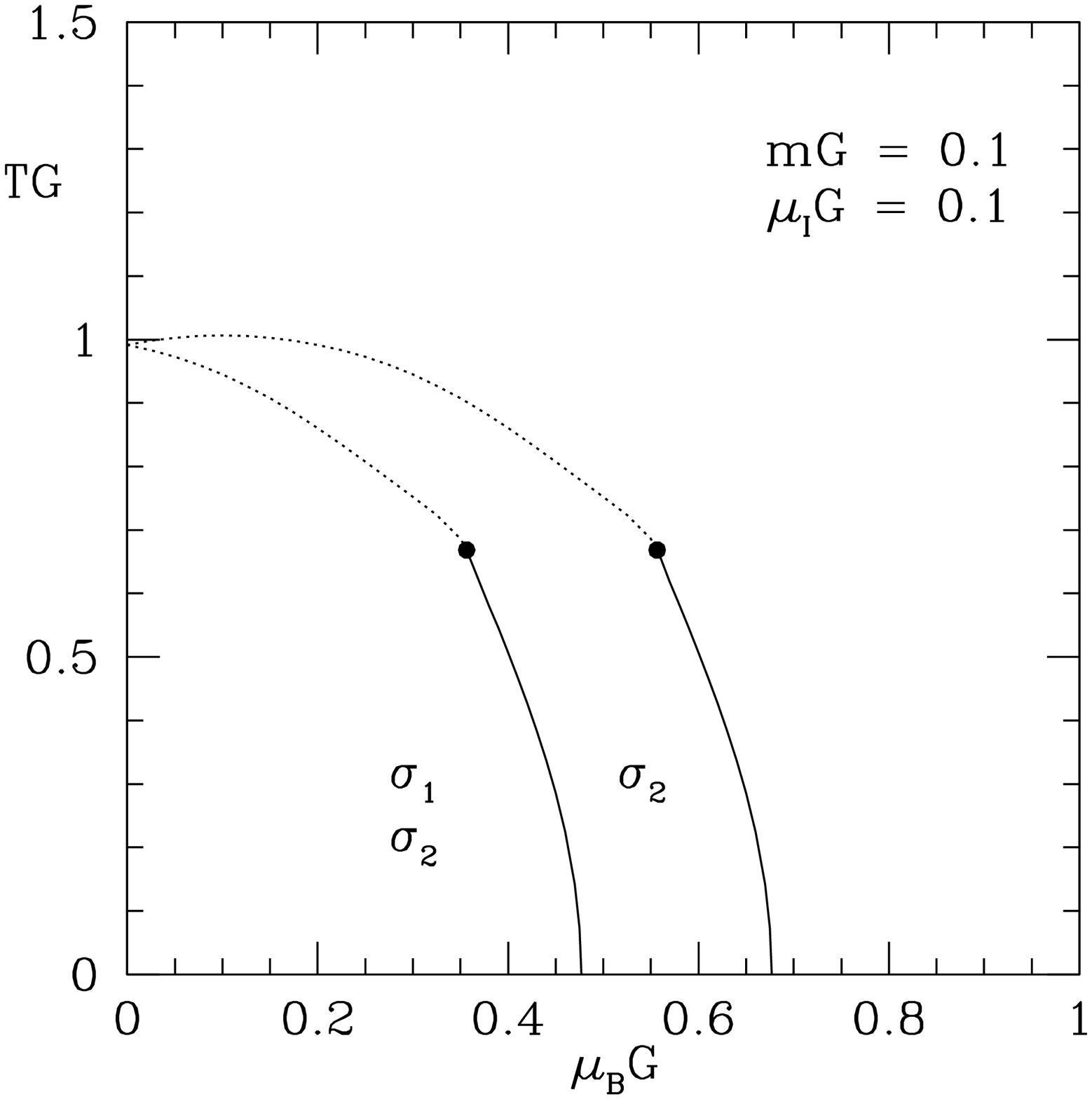}
\caption{Phase diagram in the $\mu_B$-$T$-plane for
quark mass  $mG=0.1$ and an 
isospin chemical as shown in the label of the figure.
A first order chiral restoration phase transition takes place at the full
line that ends in the critical end point. For nonzero isospin chemical
potential (lower figure) this curve is shifted in opposite directions
for the  chiral restoration transition of $\langle \bar u u \rangle$ and 
$\langle \bar d d \rangle$. The condensates that are not displayed are
of ${\mathcal O}(m)$.
The dotted curves depict the crossover behavior. The temperature of
the critical end point is not affected by the isospin chemical potential. 
\label{fig5}} 
\end{figure}
\subsection{Phase diagram at nonzero quark mass and temperature}

Finally, we study the phase diagram at nonzero quark mass and temperature. 
In this case it is no longer possible to obtain an analytical solution
of the saddle point equations. Instead, we determine the
minimum of the free energy numerically. The results for the phases in the
$\mu_1$-$\mu_2$-plane  for temperatures of $TG= 0.3$, $TG= 0.5$, $TG= 
0.6$, $TG= 0.8$ are shown in Fig.~\ref{fig4}. The quark mass used here is 
$mG =0.1$. 
Below the critical temperature, the phases are the same as at zero
temperature (see Fig.~\ref{fig3}) and differ qualitatively from the
chiral limit only in the central region (see Fig.~\ref{fig2}).

The phase diagram in the $\mu_B$-$T$-plane for $\mu_I=0$ has been studied 
in \cite{shrock}. In the chiral limit, the chiral restoration transition 
extends as a second order line from the $\mu_B=0$ axis, changes order at a 
tricritical point, and intersects the $T=0$ axis as a line of first order 
transitions.
For nonzero quark mass, the first order transition ends in a critical 
point, and the second order transition becomes a crossover. 
 
Fig.~\ref{fig5} shows the phase diagram in the $\mu_B$-$T$-plane at finite 
quark mass $mG=0.1$ for zero isospin chemical potential and for
$\mu_I G = 0.1$. We observe that the first order curve splits into
two first order curves that are separted by $2\mu_IG$. 
This can be
understood as follows.  Below 
the threshold for pion condensation,
the free energy separates into a sum over the two flavors.
For $\mu_I=0$, the chiral phase 
transition lines for both flavors coincide. A finite isospin chemical 
potential breaks the flavor symmetry, and the first order transition 
lines for the two flavors split and shift according to
\be
\mu_{B, c}^{(1)}(T) &=& \mu_c(T) -\mu_I \nn\\
\mu_{B, c}^{(2)}(T) &=& \mu_c(T) +\mu_I, 
\ee
where $\mu_c(T)$ describes the transition line at $\mu_I=0$. The critical 
temperature is not affected by the isospin chemical potential. 

We have seen in eq. (\ref{IBsymm})  that in the phase
with zero pion condensate
the dependence of the critical temperature on the isospin chemical
potential at zero baryon chemical potential is the same as the 
dependence of the critical temperature on the baryon chemical
potential at zero isospin chemical potential. This suggests the possibility
that for large values of the pion mass a tricritical point may appear in
the $\mu_B =0$ plane. As we will see below, this turns out not to be the case.
We first determine the domain in the $\mu_B =0$ plane 
where pion condensation occurs.

Performing a similar analysis as below eq. (\ref{5.41}) for the free energy 
at $\mu_B=0$ but nonzero $m$, $T$ and $\mu_I$, we find that the region
of nonzero pion condensate is bounded by the curve
\be
\mu_I^2(\mu_I^2-m^2)G^2 - \frac{1}{4}m^2 
-(\mu_I^2+T^2)(\mu_I^2-m^2)^2G^4=0.\nn\\ 
\ee
For asymptotically large values of the quark mass, this curve 
reduces to the two expressions,  
\be
\mu_{Ic}^2(T)G^2 & = & m^2 G^2 +\frac{1}{2} \pm 
\frac{1}{2mG}\sqrt{\frac{1}{2} -T^2G^2} \nn\\
&&\hspace*{-2cm} + \frac 1{8m^2G^2}(1-4T^2G^2) + ({\cal O}(\frac{1}{m^3G^3}).
\label{isodom}
\ee

However, for $\mu_B=0$ the saddle point equations are also solved by
$\rho=0$. 
Then a first order transition takes place
between the solutions with large mass expansion given by
\be
\sigma_1=\sigma_2 = 
m \pm \frac 1G \sqrt{\frac 12 - G^2T^2} + {\cal O}(\frac 1{mG}).
\ee
At the tricritical point these solutions merge with the extremum
between them. From this condition we find that the position of the
tricritical point is given by
\be
T^2_{3} G^2 &=& \frac{1}{2} -\frac{1}{16 m^2 G^2} + {\cal 
O}(\frac{1}{m^3G^3}) \\
\mu_{B, 3}^2G^2&=& m^2 G^2 + \frac{1}{2} - \frac{1}{8m^2G^2}+ {\cal 
O}(\frac{1}{m^3G^3}), \nn
\ee
and the value of $\sigma$ at the tricritical point is equal to
\be
\sigma_3 = \frac 1{4mG^2} + {\cal O}(\frac 1{m^2G^2}).
\ee 
One easily verifies from (\ref{isodom}) that up to order $1/m^2G^2$ 
the tricritical point is inside the pion condensation region, where the
phase with a nonzero pion condensate is favored.
Numerically, one finds that this is also the case for quark masses
that are not asymptotically large.

\renewcommand{\theequation}{6.\arabic{equation}}
\setcounter{equation}{0}
\section{Discussion}
\label{section-6}
Starting from  a random matrix model at nonzero temperature and chemical
potential for baryon number and isospin, we have obtained an
effective potential for the matrix valued order parameter field. 
This order parameter
field arises naturally in this random matrix model that is based on
the global symmetries of the QCD partition function.
The expectation value of its
diagonal elements are the chiral condensates
$\langle\bar u u\rangle$ and $\langle \bar dd\rangle$, whereas its
off-diagonal elements give the pion condensate. 
To first
order in $m_\pi^2$ and $ \mu_I^2$ and zero baryon chemical potential
and temperature, the random matrix model coincides with the zero
momentum part of the chiral Lagrangian that has been derived from
QCD. However, the tricritical point found in lattice simulations
\cite{dom-kog} and in the chiral Lagrangian at nonzero temperature and
isospin chemical potential \cite{STV2} is not present in the  random
matrix model. 
We therefore conclude that the pion dynamics are important for the
emergence of this tricritical point, as was suggested in
\cite{dom-kog}. 

Based on the effective potential, we have obtained a phase diagram
for QCD at nonzero temperature, baryon and isospin chemical potentials.
We have found a surprisingly rich
phase diagram characterized by the condensates in our order parameter
field. We find
that close to the critical baryon chemical potential a small isospin
chemical potential leads to a phase with $\langle \bar u u
\rangle$ of the order of $\Lambda_{QCD}^3$   but $\langle \bar d d\rangle$
reduced by a factor $m/\Lambda_{QCD}$. 

In the phase with a vanishing pion condensate, the effective potential
is an even function of the chemical potentials and separates into a
sum of free energies for each of the two flavors. This has important
consequences.
Since the effective potential is even,  
the dependence of the partition function on
$\mu_B$ at $\mu_I =0$ is the same as its dependence on $ \mu_I$ at $\mu_B = 0$.
Therefore, the phase diagram for
baryon chemical potential smaller than the pion mass can
be studied reliably by means of the phase quenched partition function.
Because of the separability of the free energy, the critical curve for
$\mu_I=0$ splits into two curves shifted by a distance of $2\mu_I$. 
As illustrated in Fig.~5, the structure of the phase
diagram in the $\mu_B$-$T$-plane is structurally altered by an arbitrarily
small nonzero isospin 
chemical potential, even for massive quarks. For a fixed
$\mu_I<m_\pi/2$, we find that there are 
two first order phase transitions at small $T$ when $\mu_B$ is
increased. Both first order lines end at the same temperature in
critical endpoints with a separation proportional to
$\mu_I$. The existence of two first order phase transition lines and
two critical endpoints  might have very important consequences
for the numerous phenomenological systems where {\it both} $\mu_B$
and $\mu_I$ are nonzero, such as neutron stars or heavy ion collision
experiments. Furthermore, it has been shown that relativistic heavy ion
collisions experiments might
be sensitive to the critical end point in the $\mu_B$-$T$-plane for
$\mu_I=0$ \cite{RHICtricPt,RHICtricPt2}. Our analysis shows that an
increase in $\mu_I$  
results in  a critical end point with a  lower value for $\mu_B$, 
thus making it easier to
reach via heavy ion collision experiments. Our analysis also implies
that two crossovers separate the quark-gluon plasma and the hadronic
phase at small but nonzero baryon and isospin chemical potentials. 
Therefore the transition between these two phases
should appear smoother at $\mu_I\neq0$ than at $\mu_I=0$.
These results have  important phenomenological consequences. It is
essential to confirm them by means of  lattice QCD simulations 
or within other models.  
\vspace*{1cm}

\begin{acknowledgments}
J. Kogut, K. Splittorff and B. Vanderheyden are acknowledged for useful
discussions. K. Splittorff is thanked for a critical reading of the
manuscript.
D. T. is supported in part by the "Holderbank"-Stiftung. This 
work was partially supported by the US DOE grant DE-FG-88ER40388 and by 
the NSF under grant NSF-PHY-0102409. 
\end{acknowledgments}


\begin{thebibliography}{99}

\bibitem{Rajagopal-Wilczek2} 
K.~Rajagopal and F.~Wilczek, 
in {\it At the Frontier of Physics/Handbook of QCD}, 
edited by M. Shifman (World Scientific, Singapore, 2001), 
Vol. 3, p. 2061.


\bibitem{Hands:2001jn}
S.~Hands,
Nucl.\ Phys.\ Proc.\ Suppl.\  {\bf 106}, 142 (2002).

\bibitem{fodor}
Z.~Fodor and S.~D.~Katz, 
JHEP {\bf 03}, 014 (2002).

\bibitem{owe}
P.~de Forcrand and O.~Philipsen, 
Nucl.\ Phys.\ {\bf B642}, 290 (2002).

\bibitem{frith}
C.~R.~Allton {\it et al.}, 
Phys.\ Rev.\ D {\bf 66}, 074507
(2002). 

\bibitem{maria}
M.~D'Elia and M.~P.~Lombardo, 
arXiv:hep-lat/0209146.

\bibitem{gupta}
S.~Gupta,
arXiv:hep-lat/0202005.

\bibitem{miya}
O.~Miyamura, S.~Choe, 
Y.~Liu, T.~Takaishi, and A.~Nakamura,
Phys.\ Rev.\ D {\bf 66}, 077502 (2002).

\bibitem{gavai}
R.~V.~Gavai and S.~Gupta,
Phys.\ Rev.\ D {\bf 65}, 094515 (2002).

\bibitem{gatto}
A.~Barducci, R.~Casalbuoni, S.~De Curtis, 
R.~Gatto and G.~Pettini,
Phys.\ Lett. {\bf B231}, 463 (1989);
Phys.\ Rev.\ D {\bf 41}, 1610 (1990).

\bibitem{shrock} M.~A.~Halasz, A.~D.~Jackson, 
R.~E.~Shrock, M.~A.~Stephanov, 
and J.~J.~M.~Verbaarschot, 
Phys.\ Rev.\ D {\bf 58} 096007 (1998). 


\bibitem{berges}
J.~Berges and K.~Rajagopal,
Nucl.\ Phys.\ {\bf B538}, 215 (1999).


\bibitem{frank}
M.~G.~Alford, A.~Kapustin, and F.~Wilczek,
Phys.\ Rev.\ D {\bf 59}, 054502 (1999).

\bibitem{kogiso}
J.~B.~Kogut and D.~K.~Sinclair,
Phys.\ Rev.\ D {\bf 66}, 014508 (2002);
Phys.\ Rev.\ D {\bf 66}, 034505 (2002).

\bibitem{kst} 
J.~B.~Kogut, M.~A.~Stephanov, 
and D.~Toublan, 
Phys.\ Lett. {\bf B464}, 183 (1999).

\bibitem{kstvz} 
J.~B.~Kogut, M.~A.~Stephanov, D.~Toublan, 
J.~J.~M.~Verbaarschot, and A.~Zhinitsky, 
Nuc.\ Phys. {\bf B582}, 477 (2000).

\bibitem{TV}
D.~Toublan and J.~J.~M.~Verbaarschot, 
Int.\ J.\ Mod.\ Phys.\ B {\bf 15}, 1404 (2001).

\bibitem{misha-son}
D.~T.~Son and M.~A.~Stephanov,
Phys.\ Rev.\ Lett. {\bf 86} (2001) 592.

\bibitem{domstrange}
J.~B.~Kogut and D.~Toublan,
Phys.\ Rev.\ D {\bf 64}, 034007 (2001).
 
\bibitem{SST}
K.~Splittorff, D.~T.~Son, and M.~A.~Stephanov,
Phys.\ Rev.\ D {\bf 64}, 016003 (2001).

\bibitem{SSSTV}
T.~Sch{\"a}fer, D.~T.~Son, 
M.~A.~Stephanov, D.~Toublan, 
and J.~J.~M.~Verbaarschot, 
Phys.\ Lett.\ {\bf B522}, 67 (2001).


\bibitem{sanvec}
F.~Sannino and W.~Sch{\"a}fer,
Phys.\ Lett.\ {\bf B527}, 142 (2002).


\bibitem{STV} 
K.~Splittorf, D.~Toublan, 
and J.~J.~M.~Verbaarschot, 
Nucl.\ Phys. {\bf B620}, 290 (2002).

\bibitem{kimvec}
K.~Splittorff,
arXiv:hep-lat/0110226.

\bibitem{STV2}
K.~Splittorff, D.~Toublan, 
and J.~J.~M.~Verbaarschot,
Nucl.\ Phys. {\bf B639}, 524 (2002).



\bibitem{shins}
G.~V.~Dunne and S.~M.~Nishigaki,
arXiv:hep-ph/0210219.

\bibitem{lat1}
S.~Hands, I.~Montvay, S.~Morrison, 
M.~Oevers, L.~Scorzato, and 
J.~Skullerud,
Eur.\ Phys.\ J.\ C {\bf 17}, 285 (2000).

\bibitem{lat2}
R.~Aloisio, A.~Galante, V.~Azcoiti, 
G.~Di Carlo, and A.~F.~Grillo,
arXiv:hep-lat/0007018.

\bibitem{lat3}
R.~Aloisio, V.~Azcoiti, G.~Di Carlo, 
A.~Galante, and A.~F.~Grillo,
Phys.\ Lett.\ {\bf B493}, 189 (2000).

\bibitem{lat4}
Y.~Liu, O.~Miyamura, A.~Nakamura and T.~Takaishi,
arXiv:hep-lat/0009009.


\bibitem{lat5}
S.~J.~Hands, B.~Kogut, 
S.~E.~Morrison, and D.~K.~Sinclair,
Nucl.\ Phys.\ Proc.\ Suppl.\  {\bf 94}, 457 (2001).



\bibitem{lat6}
R.~Aloisio, V.~Azcoiti, G.~Di Carlo, 
A.~Galante, and A.~F.~Grillo,
Nucl.\ Phys.\ {\bf B606}, 322 (2001).

\bibitem{lat7}
J.~B.~Kogut, D.~K.~Sinclair, 
S.~J.~Hands, and S.~E.~Morrison,
Phys.\ Rev.\ D {\bf 64}, 094505 (2001).



\bibitem{dom-kog}
J.~B.~Kogut, D.~Toublan, and D.~K.~Sinclair, 
Phys.\ Lett. {\bf B514}, 77 (2001);
Nucl.\ Phys.\ {\bf B642}, 181 (2002).

\bibitem{Bailin:bm}
D.~Bailin and A.~Love,
Phys.\ Rept.\  {\bf 107}, 325 (1984).

\bibitem{Son:1998uk}
D.~T.~Son,
Phys.\ Rev.\ D {\bf 59}, 094019 (1999).

\bibitem{krishna}
M.~G.~Alford, K.~Rajagopal, 
and F.~Wilczek,
Phys.\ Lett.\ {\bf B422}, 247 (1998).



\bibitem{edward}
R.~Rapp, T.~Sch{\"a}fer, 
E.~V.~Shuryak, and M.~Velkovsky,
Phys.\ Rev.\ Lett.\  {\bf 81}, 53 (1998).

\bibitem{SV} 
E.~V.~Shuryak and J.~J.~M.~Verbaarschot,
Nucl.\ Phys.\ {\bf A560}, 306 (1993).

\bibitem{V} 
J.~J.~M.~Verbaarschot,
Phys.\ Rev.\ Lett. {\bf 72}, 2531 (1994).

\bibitem{OTV} 
J.~C.~Osborn, D.~Toublan,
and J.~J.~M.~Verbaarschot, 
Nucl.\ Phys. {\bf B540}, 317 (1999). 

\bibitem{DOTV} 
P.~H.~Damgaard, J.~C.~Osborn, D.~Toublan, 
and J.~J.~M.~Verbaarschot,  
Nucl. Phys. {\bf B547}, 305 (1999).   

\bibitem{TV-correl}
D.~Toublan and J.~J.~M.~Verbaarschot,
Nucl.\ Phys.\ {\bf B603}, 343 (2001).

\bibitem{TV-beta}
D.~Toublan and J.~J.~M.~Verbaarschot,
Nucl.\ Phys.\ {\bf B560}, 259 (1999).


\bibitem{JV} 
A.~D.~Jackson and J.~J.~M.~Verbaarschot, 
Phys.\ Rev.\ D {\bf 53},  7223 (1996). 

\bibitem{misha} 
M.~A.~Stephanov, 
Phys.\ Rev.\ Lett. {\bf 76}, 4472 (1996).

\bibitem{adam}
M.~A.~Halasz, J.~C.~Osborn, M.~A.~Stephanov,
and J.~J.~M.~Verbaarschot,
Phys.\ Rev.\ D {\bf 61}, 076005 (2000).

\bibitem{jun}
J.~Ambjorn, K.~N.~Anagnostopoulos, 
J.~Nishimura, and J.~J.~M.~Verbaarschot,
arXiv:hep-lat/0208025.

\bibitem{Benoit1} 
B.~Vanderheyden and A.~D.~Jackson, 
Phys.\ Rev.\ D\ {\bf 62}, 094010 (2000).

\bibitem{Benoit2} 
B.~Vanderheyden and A.~D.~Jackson, 
Phys.\ Rev.\ D\ {\bf 61}, 076004 (2000).

\bibitem{Benoit3} 
B. Vanderheyden and A.~D.~Jackson, 
Phys.\ Rev.\ D\ {\bf 64},  074016 (2001).

\bibitem{Pepin} 
S.~Pepin and A.~Sch\"afer, 
Eur.\ Phys.\ J.\ A\ {\bf 10}, 303 (2001).
 
\bibitem{benoitspect}
B.~Vanderheyden and A.~D.~Jackson,
arXiv:hep-ph/0208085.


\bibitem{RHICtricPt}
M.~A.~Stephanov, K.~Rajagopal, 
and E.~V.~Shuryak,
Phys.\ Rev.\ Lett.\  {\bf 81}, 4816 (1998); 
Phys.\ Rev.\ D {\bf 60}, 114028 (1999).

\bibitem{RHICtricPt2}
Y.~Hatta and T.~Ikeda,
arXiv:hep-ph/0210284.

\bibitem{Hands:2001ck}
S.~Hands,
Nucl.\ Phys.\ {\bf A702}, 206 (2002).


\end{thebibliography}
\end{document}